\input{aipcheck}
\documentclass[final]{aipproc}
\layoutstyle{6x9}

\begin{document}

\title{Finding Supernova Ia Progenitors with the Chandra X-ray Observatory}

\classification{97, 98}
\keywords      {binaries: close - supernovae: general - X-ray: binaries}

\author{Mikkel Nielsen}{address={Department of Astrophysics, Radboud University Nijmegen, the Netherlands}}
\author{Gijs Nelemans}{address={Department of Astrophysics, Radboud University Nijmegen, the Netherlands}}
\author{Rasmus Voss}{address={Department of Astrophysics, Radboud University Nijmegen, the Netherlands}}

\begin{abstract}
We examine pre-supernova Chandra images to find X-ray luminosities of type Ia supernova progenitors. At present, we have one possible direct detection and upper limits for the X-ray luminosities of a number of other supernova progenitors. The method has also yielded a possible detection of a X-ray binary Wolf-Rayet system as the progenitor of a type Ib supernova.
\end{abstract}

\maketitle


Supernovae (SNe) Ia are highly luminous stellar explosions that can be used as cosmological distance indicators. They are also the main source of iron in the Universe. As such, they are an important phenomenon in both cosmology and galactic astrophysics. However, the exact nature of their progenitor systems remains elusive. SNe Ia are believed to arise when white dwarfs (WDs) accrete matter and approach the Chandrasekhar mass. At around 1.37 M$_{\odot}$ they explode in a thermonuclear runaway as carbon is rapidly fused into heavier elements. 

Currently, two progenitor scenarios are considered: the single degenerate (SD) and double degenerate (DD) scenarios. In the SD scenario, a WD accretes mass from a companion star and eventually explodes. In the DD scenario, two sub-Chandrasekhar mass WDs merge to form a single WD above the Chandrasekhar mass which subsequently explodes. Both scenarios face considerable problems, however, and a full understanding of the properties of SNe Ia progenitors is sorely needed \cite{DiStefano.2010}.

A sufficiently massive WD accreting mass from a companion star will burn accreted material at its surface. The energy produced in this process has been shown to dominate over the acceleration of the infalling mass (unlike for more compact X-ray accretors, i.e. neutron stars or black holes, where the acceleration of matter dominates) and will produce soft X-rays at luminosities around $\sim 10^{37}-10^{38}$ erg/s. DD progenitors likely have no or weaker X-ray emission. Therfore, we examine pre-explosion archival images from the Chandra satellite to find the X-ray brightness of the progenitors and thereby determine whether they are SD or not.

By this approach Voss \& Nelemans \cite{Voss.Nelemans.2008} found evidence of an X-ray emitting progenitor of SN2007on. However, later analyses indicate what might have been a chance alignment between the progenitor and an unidentified X-ray source \cite{Roelofs.et.al.2008}. At present, it is unclear whether the X-ray source in question was in fact the progenitor of SN2007on. More SNe need to be studied.

\begin{table}[!ht]
    \begin{tabular}{c c}
      \hline
      SN & L$_X$ \\
      \hline
      2008fp & $1.5\cdot10^{38}$ erg/s \\
      2007sr & $1.9\cdot10^{37}$ erg/s \\
      2007on & $5.1\cdot10^{38}$ erg/s \\
      2007gi & $3.6\cdot10^{38}$ erg/s \\  
      2006mr & $4.8\cdot10^{38}$ erg/s \\  
      2006dd & $3.1\cdot10^{38}$ erg/s \\  
      2006X  & $2.5\cdot10^{39}$ erg/s \\
      2004W  & $4.0\cdot10^{37}$ erg/s \\
      2003cg & $1.6\cdot10^{39}$ erg/s \\
      2002cv & $1.9\cdot10^{38}$ erg/s \\ \hline
    \end{tabular}
  \caption{3$\sigma$ upper limits on X-ray luminosities of SN Ia progenitors} \label{upper.lims}
\end{table}

Pre-SN Chandra images of nearby SNe Ia allow upper limits to be calculated for X-ray luminosities of progenitors (see table \ref{upper.lims}). We note that these are close to theoretical predictions for accreting white dwarfs. More nearby SNe Ia are likely to further constrain these values and/or provide a direct detection of a progenitor.

To generate our sample we use the list of SNe provided by the Harvard CfA (\url{http://www.cfa.harvard.edu/iau/lists/Supernovae.html}). Chandra became operational in October 1999 so we only consider SNe later than this time. Since luminosities of $\sim10^{37}$ erg/s are unlikely to be detectable at larger distances than 25 Mpc we use NED (\url{http://nedwww.ipac.caltech.edu/}) to give us only those SNe Ia whose host galaxies are within this distance.

\begin{figure}[!t]
  \includegraphics[height=0.5\linewidth]{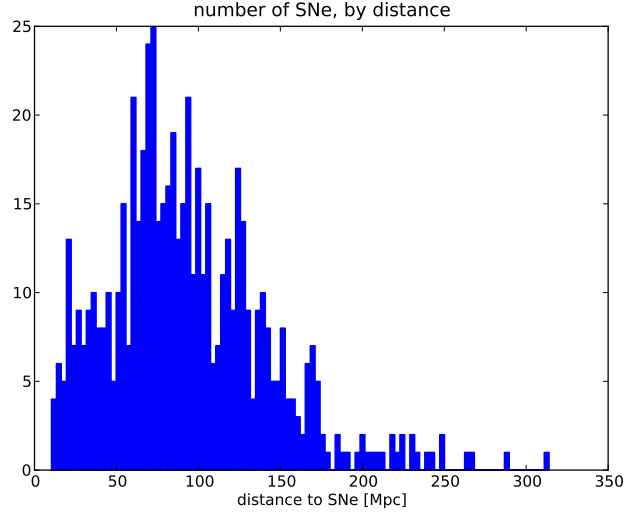}
  \centering
  \caption{SNe Ia in host galaxies at known distances.}
  \label{dist.histo}
\end{figure}

\begin{figure}[!t]
  \includegraphics[height=0.5\linewidth]{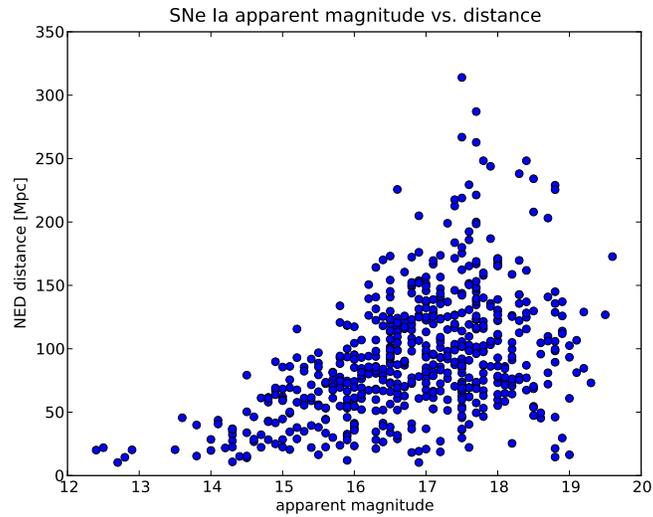}
  \caption{Distance to nearby ($d\leq 25 Mpc$) SNe Ia as a function of apparent magnitude. We note the large scatter in apparent magnitudes, from 12.4 as the brightest (SN2006ce) to 19.0 as the dimmest (SN2002cv). Nearby SNe are often discovered by amateur astronomers whose telescopes are hard pressed to detect anything below magnitude 17 or so, hence we expect a significant number of dim, undiscovered nearby SNe. This raises serious concerns with respect to the completeness of our sample.}
  \label{dist.mag}
\end{figure}

\begin{figure}[!ht]
  \centering
  \includegraphics[height=0.75\textheight]{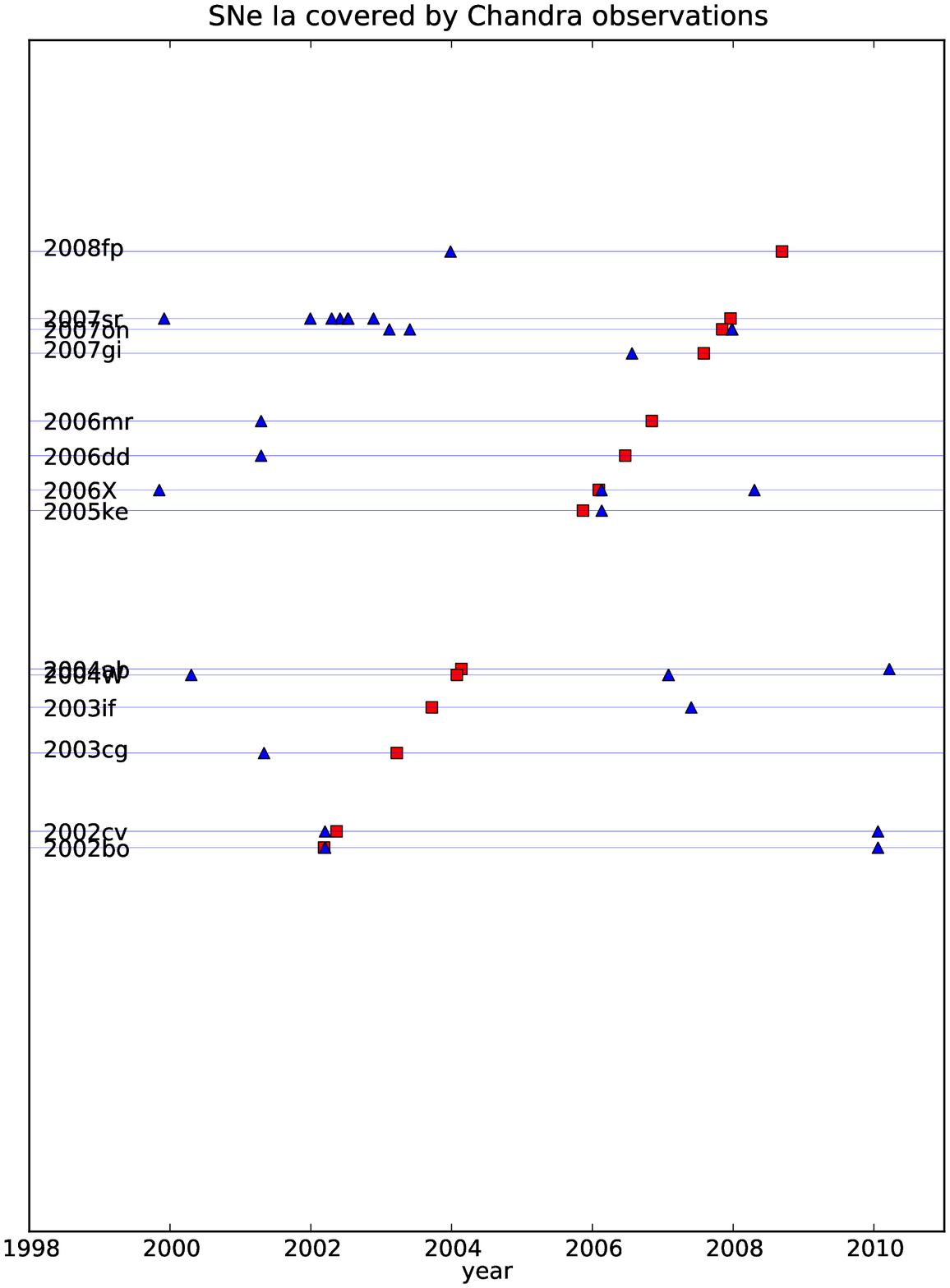}
  \caption{Timeline of SNe Ia covered by Chandra. Each horizontal line on the plot corresponds to a nearby SN for which Chandra data is available: 2002bo, 2002cv, 2003cg, 2003if, 2004W, 2004ab, 2005ke, 2006X, 2006dd, 2006mr, 2007gi, 2007on, 2007sr, \& 2008fp. Squares mark the date of SNe Ia while triangles mark observations covering the position of the SN. Hence, triangles to the left of the square on the same line indicate pre-SN images. As can be seen, there is a roughly constant distribution of observations. We note that there is a curious absence of Chandra-covered SNe from 1999 to 2002. We would expect there to be at least a couple of SNe with post-SN images in this region.}
  \label{obsdates.SNdates}
\end{figure}

The total number of SNe Ia with known distances is 608. Figure \ref{dist.histo} plots the number of SNe Ia as a function of NED distance and for a cut-off distance of 25 Mpc we arrive at a sample of 34 SNe Ia. Of this sample, we find 10 with pre-SN images (figure \ref{obsdates.SNdates}). This corresponds to approximately one SN Ia having pre-SN images per year, although this number is obviously subject to large uncertainties due to small number statistics. Completeness is also an issue of some concern, see figure \ref{dist.mag}.

In our work, we have also examined Chandra archival images for evidence of progenitors of other types of SNe (Ib/c, II). Evidence for an X-ray source was found for SN2010O, and this has been interpreted as an X-ray binary in a Wolf-Rayet system \cite{Nelemans.et.al.2010}.

Our conclusion is that the method works and have already produced a possible direct detection in addition to interesting upper limits to X-ray luminosities of the progenitors of nine SNe Ia. The method may prove helpful in the field of core-collapse SNe as well. As time progresses we expect more SNe Ia to occur that have pre-SN images available and it appears likely that this will happen soon.

\bibliographystyle{aipproc}   
\bibliography{mnielsen}

\begin{thebibliography}{4}
\expandafter\ifx\csname natexlab\endcsname\relax\def\natexlab#1{#1}\fi
\providecommand{\enquote}[1]{``#1''}
\expandafter\ifx\csname url\endcsname\relax
  \def\url#1{\texttt{#1}}\fi
\expandafter\ifx\csname urlprefix\endcsname\relax\def\urlprefix{URL }\fi
\providecommand{\eprint}[2][]{\url{#2}}

\bibitem[Di~Stefano(2010)]{DiStefano.2010}
R.~Di~Stefano, \emph{ApJ} \textbf{712}, 728--733 (2010).

\bibitem[{Voss} and {Nelemans}(2008)]{Voss.Nelemans.2008}
R.~{Voss}, and G.~{Nelemans}, \emph{Nature} \textbf{451}, 802--804 (2008).

\bibitem[{Roelofs} et~al.(2008)]{Roelofs.et.al.2008}
G.~{Roelofs}, C.~{Bassa}, R.~{Voss}, and G.~{Nelemans}, \emph{MNRAS}
  \textbf{391}, 290--296 (2008).

\bibitem[{Nelemans} et~al.(2010)]{Nelemans.et.al.2010}
G.~{Nelemans}, R.~{Voss}, M.~T.~B. {Nielsen}, and G.~{Roelofs}, \emph{MNRAS}
  \textbf{405}, L71--L75 (2010).

\end{thebibliography}

\end{document}